\documentclass[conference]{IEEEtran}
\usepackage{blindtext, graphicx}
\usepackage[autostyle=false, style=english]{csquotes}
\MakeOuterQuote{"}

\usepackage[utf8]{inputenc}
 
\usepackage{listings}
\usepackage{color}
\usepackage[utf8]{inputenc}
\usepackage[table]{xcolor}
\usepackage{wrapfig}
\usepackage{lipsum}
\usepackage{float} 
\usepackage{tfrupee} 
\usepackage[utf8]{inputenc}
 
\usepackage{algorithm}
\usepackage[noend]{algpseudocode}
\definecolor{codegreen}{rgb}{0,0.6,0}
\definecolor{codegray}{rgb}{0.5,0.5,0.5}
\definecolor{codepurple}{rgb}{0.58,0,0.82}
\definecolor{backcolour}{rgb}{0.95,0.95,0.92}
 
\lstdefinestyle{mystyle}{
    backgroundcolor=\color{backcolour},   
    commentstyle=\color{codegreen},
    keywordstyle=\color{magenta},
    numberstyle=\tiny\color{codegray},
    stringstyle=\color{codepurple},
    basicstyle=\footnotesize,
    breakatwhitespace=false,         
    breaklines=true,                 
    captionpos=b,                    
    keepspaces=true,                 
    numbers=left,                    
    numbersep=5pt,                  
    showspaces=false,                
    showstringspaces=false,
    showtabs=false,                  
    tabsize=2
}
 
\usepackage{cite}

\ifCLASSINFOpdf
\else
\fi
\usepackage[cmex10]{amsmath}
\begin{document}
%
\title{Designing a Binary Clock using logic gates}

\author{\IEEEauthorblockN{Jacob John\IEEEauthorrefmark{1}}
\IEEEauthorblockA{School of Computer Science and Engineering (SCOPE)\\
Vellore Institute of Technology, Vellore\\
Email: \IEEEauthorrefmark{1}jacob.john2016@vitalum.ac.in}}
\maketitle


%


\maketitle

\begin{abstract}
Wrist watches have been a common fashion accessory addition for several people. However, the concept of using a seven-segment digital display or sometimes, even an analog indicator hasn't changed for a number of years. This project aims to test and design a binary clock, also referred to as \textit{32, 16,  8,  4,  2,  1} clocks or even \textit{8,  4,  2,  1} clocks (due to their display configuration),  that could change this everlasting display for watches. Specifically, digital logic and design engineers would find interest in this topic due to sophistication involved in reading-out the time. This project will do so using a by showing each decimal digit of sexagesimal time as a binary value. This design will be primarily functioning on logic gates and would involve the use of several basic components that include, but are not limited to, integrated circuits (or ICs), Light emitting diodes (LEDs) and resistors. 
\end{abstract}



%
\IEEEpeerreviewmaketitle

\section{Introduction}

\subsection{Standard Seven Segment Display}

A seven-segment display, in digital electronics, is a form of electronic display device used as a means to represent a decimal digit using a seven individually lit segments. The image below shows a standard seven segment display without the decimal point:

\begin{figure}[H]
  \centering
  \includegraphics[width=3cm]{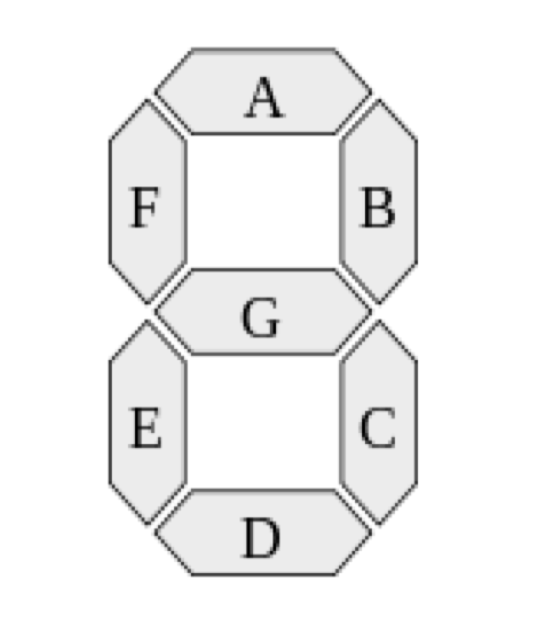}
  \caption{A standard seven segment display without decimal point \cite{Morris}}
\end{figure}

As shown in the figure, the segments of the display are illustrated and referred by the \textit{A,B,C,D,E,F and G}. An engineer can choose to select for which set of inputs will each segment LED will light up for using these characterizing letters,  \textit{A-G}.

Using this concept, a binary clock or a "complete" clock can be represented using 3 seven segment displays, 1 two-decimal point/colon display and 1 two segment display. Time uses a sexagesimal concept which can also be referred to as the base 60 concept in digital technology. This is because the minutes and seconds part of a clock only ranges from 0-59, while repeating the cycle and incrementing the preceding place by a value of 1. The first digit of the hours place can be represented by a two segment display as the most significant digit maximum value of hours \textit{(i.e. 12)} is 1. Furthermore, since the minutes place can range from \textit{0-59}, 2 seven segments displays are required.  Finally the colon or the two-decimal point display is used as a minutes to seconds separator. Hence, the final outcome for the display is as follows:

\begin{figure}[H]
  \centering
  \includegraphics[width=4cm]{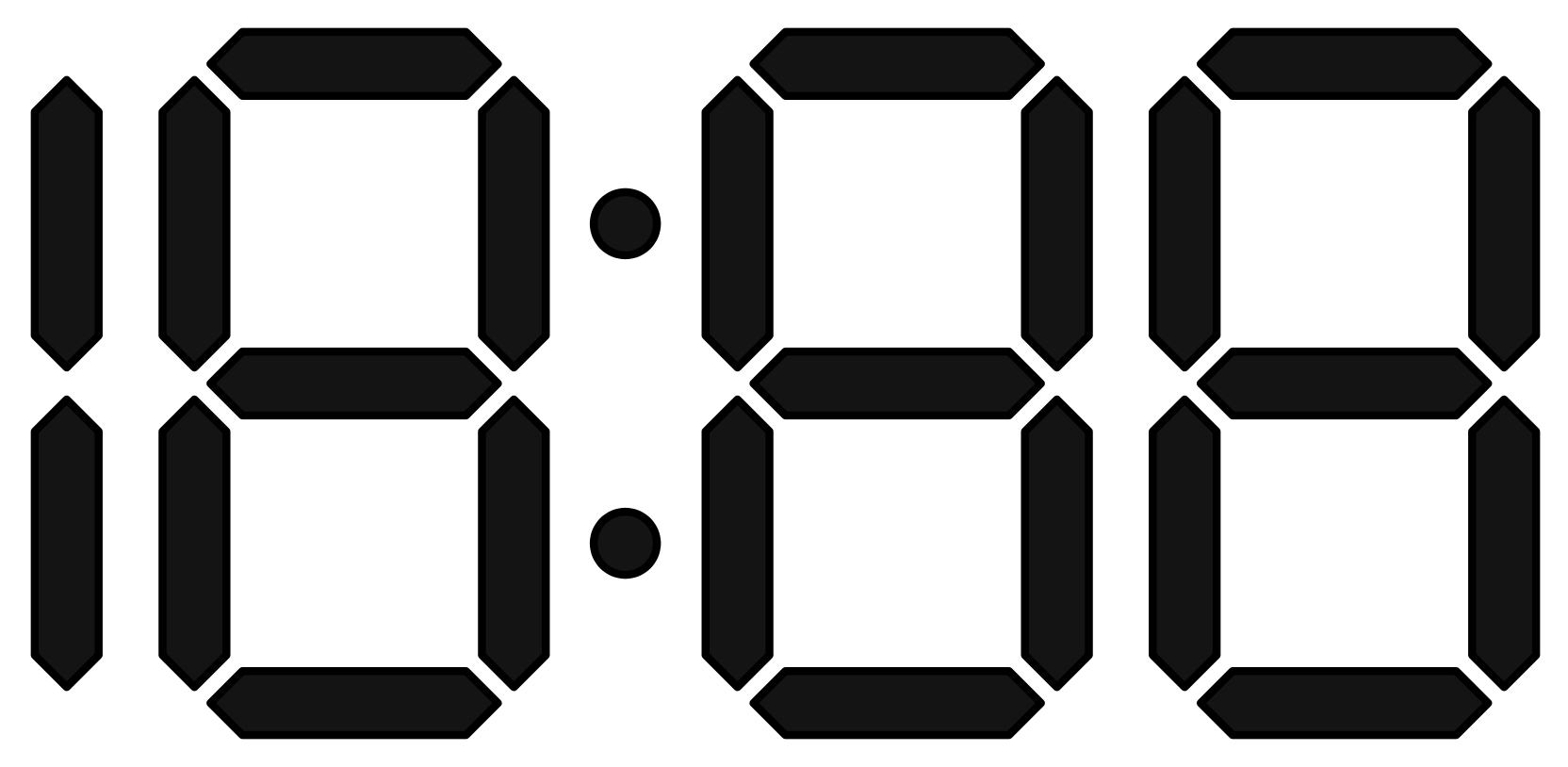}
  \caption{A standard seven segment display of a 12-hour digital clock}
\end{figure}

\underline{\textit{Note}}: Although, The circuit designed in this project will utilize a hours place, a minutes place and a seconds place for this project, it is not displayed in the figure above. This is to allow for a less complicated explanation for the principle of the seven segment displays used in 12 hour digital clocks. Furthermore, majority of 12 hour digital clocks only employ a hours place and a minutes place.

\subsection{Binary Clock Display}

A binary clock displays the time of day in a binary format. It uses a similar concept of HH:MM:SS or an hours place, minutes place and a seconds place like standard 12 or 24 hour clocks usually do. However, each decimal digit of sexagesimal or base 60 time, will be represented as a binary value. Similar clocks based on gray code and BCD logic also exist,  but our main focus is on the use of binary values. 

\begin{figure}[H]
  \centering
  \includegraphics[width=\columnwidth]{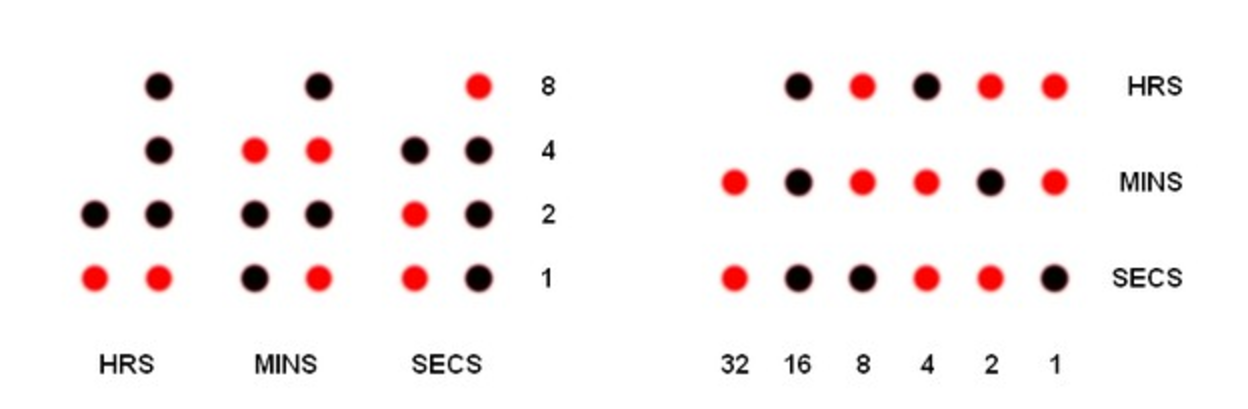}
  \caption{BCD on the left and traditional binary (as used in this clock) on the right displaying 11:45:38 in 12 hour format\cite{Stuart}}
\end{figure}

The binary clock in this project will be utilizing the traditional binary format (on the left image of Fig. 3). Time for seconds, hours and minutes are calculated by adding up the lit up LEDs to the corresponding power of 2.  On the illustration on the right,  in the circuit for the seconds part the LEDs for $ 2^{5} $, $ 2^{2} $ and $ 2^{1} $ are lit up. Hence, the seconds part is $ 32+4+2 = 38 $. For the minutes part, the LEDs for $ 2^{5} $, $ 2^{3} $, $ 2^{2} $ and $ 2^{0}$ are lit up. Hence, the output is $32+8+4+1 = 45$. Finally, for the hours section, the LEDs for $ 2^{3} $, $ 2^{1} $ and $ 2^{0} $ are lit up. Hence, the output is $8+2+1 = 11$. Using this process we obtain 11:45:38. This traditional clock is also a form of a 24 hour clock. 

The circuit on the left follows a different approach,  rather than adding digits using powers of 2 in a horizontal fashion, we take the sum in a vertical fashion. Each segment represents two parts–first the tens part and then the ones part. The first segment is the hours segment. In the figure, the tens part of the hours segment is one. This is because the LED corresponding to only 1 is lit up. Similarly, the ones part is also 1. Thus the hours segment corresponds to 11.  Using the same logic as discussed for the hours segment, the tens part of the minutes segment is 4. However, in the ones part of the minutes segment since 2 LEDs are lit up, we need to add these two numbers, i.e.,$4+1=5$. Hereby forming the number 45. Finally, the seconds segment represents 38, thus we obtain the time 11:45:38. This is the representation we will be proceeding with.  

\section{Investigation}

\subsection{Circuit Diagram}

The following circuit is a schematic of the logic diagram for the binary clock which is going to be implemented.
\begin{figure}[H]
  \centering
  \includegraphics[width=\columnwidth]{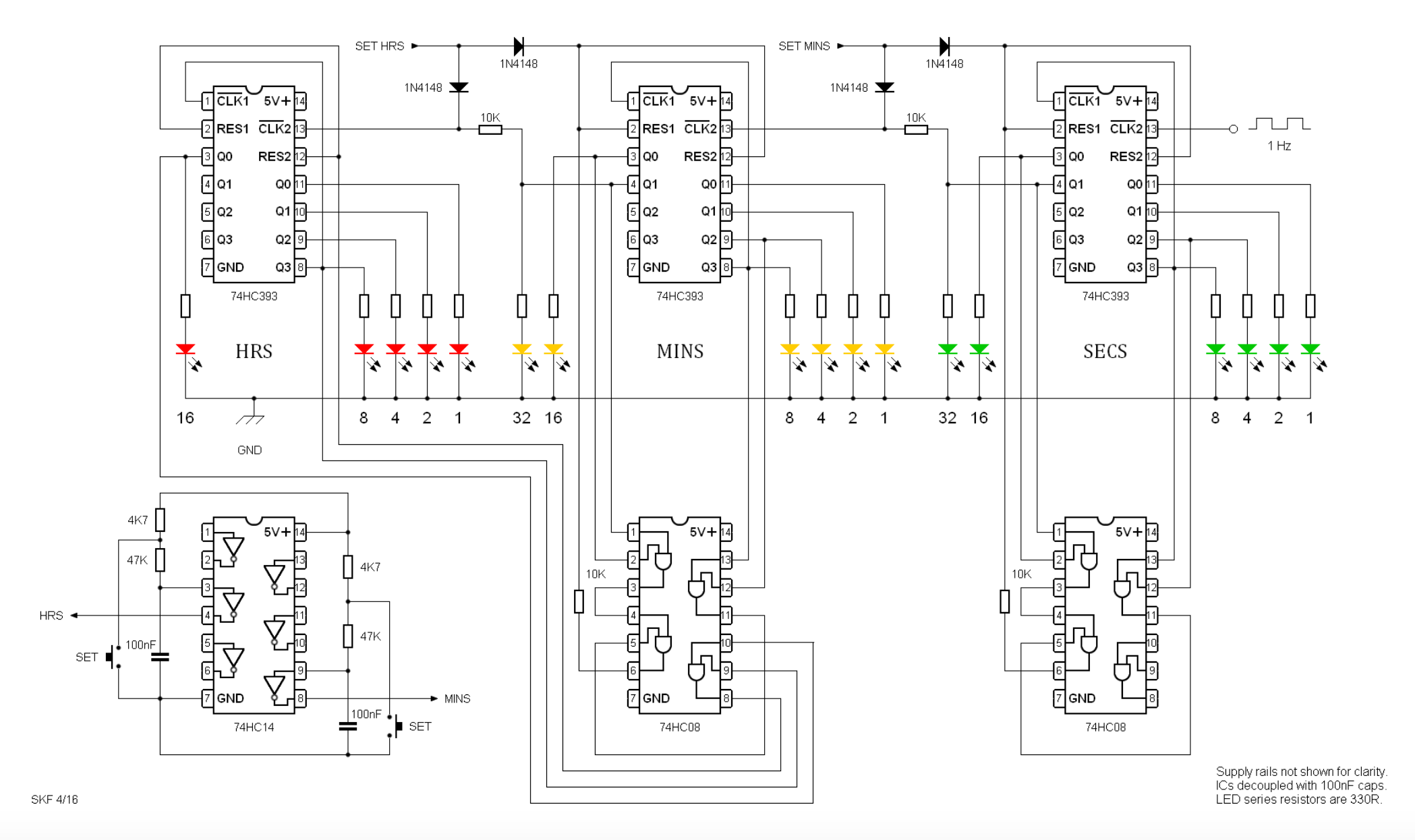}
  \caption{Logic circuit using logic gates for a binary clock \cite{Stuart}}
\end{figure}

\subsection{Components Required}

The components required for this circuit are as follows:
\begin{lstlisting}[frame=bt,numbers=none,mathescape=true]
3 $\times$ 74HC393 IC
2 $\times$ 74HC08 IC
1 $\times$ 74HC14 IC
4 $\times$ 1N4148 IC
2 $\times$ Switch (Button), for setting and resetting the time
2 $\times$ 100 nF Capacitor
4 $\times$ 47 k$\Omega$ Resistors
17 $\times$ 330 $\Omega$ Resistors
3 $\times$ 10 k$\Omega$ Resistors
1 $\times$ Arduino chip	
\end{lstlisting}

\subsection{Feasibility Study}

All the resistors, IC, Clock crystal, capacitor, LEDs and bread board are also available locally. So all the required parts can be acquired easily and are within our scope. The prices in the table below are in \textit{Indian Rupees} (INR or \rupee).

\begin{tabular}{ |p{4.1cm}||p{1.25cm}|p{1cm}|  }
 \hline
 \multicolumn{3}{|c|}{\textbf{Estimated Costs}} \\
 \hline
Component Name & Quantity & Cost\\
 \hline
 Bread Board & 2 & \rupee 200\\
 Battery & 1 & \rupee 100\\
 Resistors & 20 & \rupee 100\\
 ICs & 6 & \rupee 100\\
 Capacitors & 2  & \rupee 100\\
 LEDs & 20 & \rupee 100\\
 Arduino chip & 1 & \rupee 500\\
 \hline
 \multicolumn{2}{|l|}{\textbf{Grand Total}} & \rupee 1200 \\
 \hline
\end{tabular}

\underline{\textit{Note}}: The costs are highest bound estimates that include extra components as well. 

\underline{\textit{Note}}: IN\rupee 1200 is approximately US\$20 as of May, 2017.

\section{Methodology}

\subsection{Initial preparation}
Traditionally, time is displayed in a 24 hour format on digital clocks as mentioned earlier. Furthermore, constructing a counter that display a 24 hour format is easy as the clock always resets `0' on the 24th count, also common referred to as midnight. Meanwhile, 12 hour clocks resets to `1' on the 13th count or 1 o'clock. When designing a basic counter using registers, such as a 3-bit or 4-bit asynchronous up and/or down counters, the count traditionally resets back to 0. This is because JK edge triggered flip flops that are sensitive to pulse triggering and thus providing a continuous clock pulse to a asynchronous circuit is a very tedious task. Thus, making the counter design far more complicated for a 12 hour clock. 

This project is going to be using logic gates and binary ripple counters to make a 24 hour binary clock due to the difficulty we may face while constructing a 12 hour binary clock. As shown in Fig. 3, the final binary clock should display time in binary format, which will represent time as 24 hour digital clock would, i.e., HH:MM:SS. 

\subsection{Counting up}

The display section of the 12 hour clock utilizes three 74HC393 ICs. 74HC393 is a dual 4-bit binary ripple counter that contains two 4-bit binary counters with individual clocks. Furthermore, it also contains two master resets to clear each 4-bit counter
individually. \cite{Philips} This allows for resetting of the counter once the final count is reached. A circuit diagram for this IC is given below:

\begin{figure}[H]
  \centering
  \includegraphics[width=4cm]{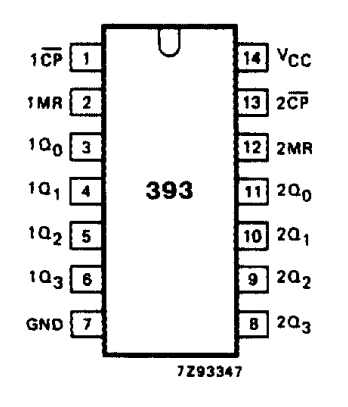}
  \caption{74HC393: Dual 4-bit binary ripple counter \cite{Philips}}
\end{figure}

The two 4-bit binary counters can be used to produce a single a single 8-bit counter. For this project however, only 6 bits are needed for minutes and for seconds while 5 bits are required for the hours part. Furthermore, two AND gates or 74HC08 quad AND gate ICs are used to reset the counters. Resetting will take place once the desired binary number has reached it's highest value. In the case of minutes and for seconds, the resetting should take place at `60', so that `60' is not actually displayed on the clock. For hours, on the other hand, resetting should place at `24'. The schematic this project is using is displayed as Fig. 5 of this report.

1 Hz pulses are fed into the clock input of the seconds 74HC393 IC as shown in Fig. 7. The counters are triggered by a HIGH-to-LOW transition of the clock inputs. \cite{Philips} This is also known as negative edge triggering in the case of a flip flop, specifically, known as a register as it is sensitive to pulse transitions. The reason for choosing a 1 Hz frequency for the clock pulse is because the counter would increment by 1 every time the clock pulse transitions, which is every second. Thus making it suitable for a binary \textit{clock}.

\begin{figure}[H]
  \centering
  \includegraphics[width=5cm]{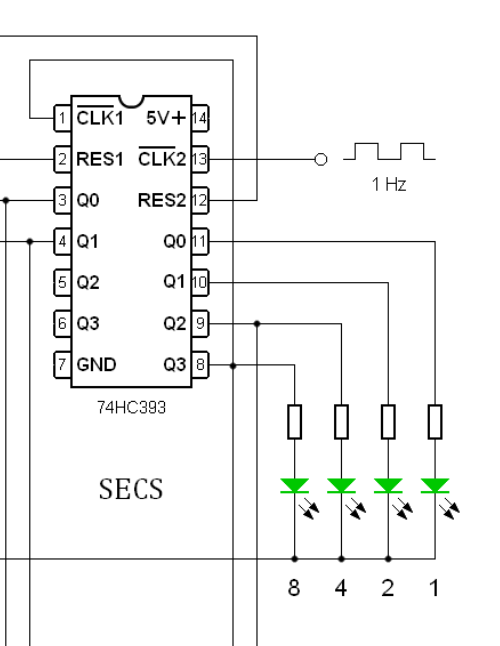}
  \caption{74HC393 seconds counter being fed with a 1Hz clock pulse \cite{Stuart}}
\end{figure}

The counter above is the seconds counter which counts up to `59' and then resets back to `0' on the 60th count. This can be done with the use of an AND gate and a simple combinational logic. Since the maximum count is produced when the four pins- 8, 9, 3 and 4 are active or when $ 4+8+16+32 = 60 $, the output of the AND gate will only be HIGH when these four pins have a HIGH output. This will send a LOW to HIGH signal to the \textit{RES1} of the counter, which will clear or reset the counter. On the other hand, the clock inputs require a transition from HIGH to LOW to the advance the clock.

\begin{figure}[H]
  \centering
  \includegraphics[width=8cm]{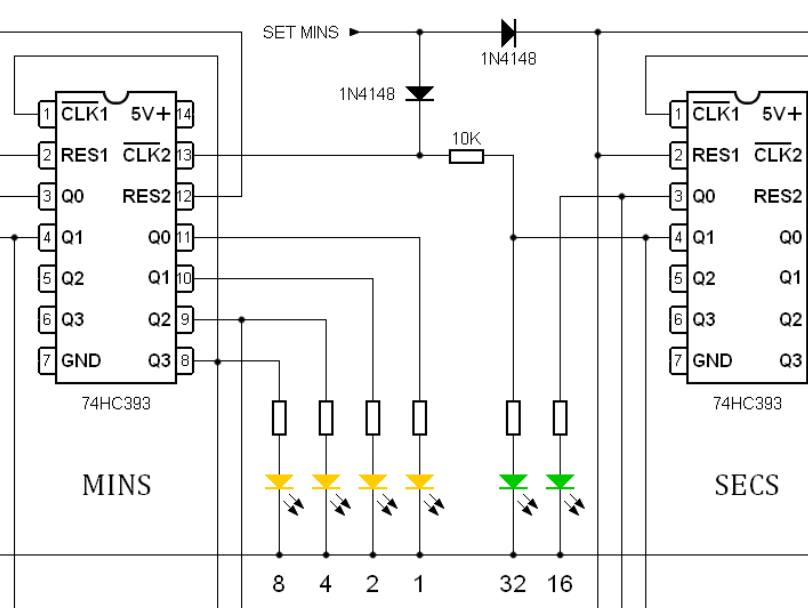}
  \caption{The most significant ouput of the Seconds counter being connected to the clock of the minutes counter\cite{Stuart}}
\end{figure}

The negative edge triggering of the clock input be used to our advantage. The most significant output, i.e. 32 will be connected to the clock count of the next counter. This means that on the count of 32, the output will be HIGH. But nothing will happen yet as even the AND gate is waiting for the inputs connected to it to become HIGH. Once they do, the counter resets and all the outputs connected to it, including 32, becomes LOW. This will create a negative edge trigger and the clock will go to the next stage. The same way, the minute also triggers the hours clock. When the time reaches 23:59:59 the next clock pulse will reset the whole clock to zero and for one second at midnight the display will be completely blank. 

\subsection{Setting the clock}

In this clock, only two push buttons are used for setting the time instead of three. These two push buttons will be used to set the hours, minutes and seconds part of the binary clock. The hours are set initially. Once the hours are set, the minutes part will be zeroed. Next, the minutes part will be set. Once the minutes part is set, the seconds part will be zeroed. 

In order to try and set the clock as accurately as possibly, one should adopt an unconventional approach. Set the time of one minute before when you want the clock to start from. Then wait for the signal and then press the set button for the minutes part. This will start the clock with the seconds counting from zero.

\begin{figure}[H]
  \centering
  \includegraphics[width=9cm]{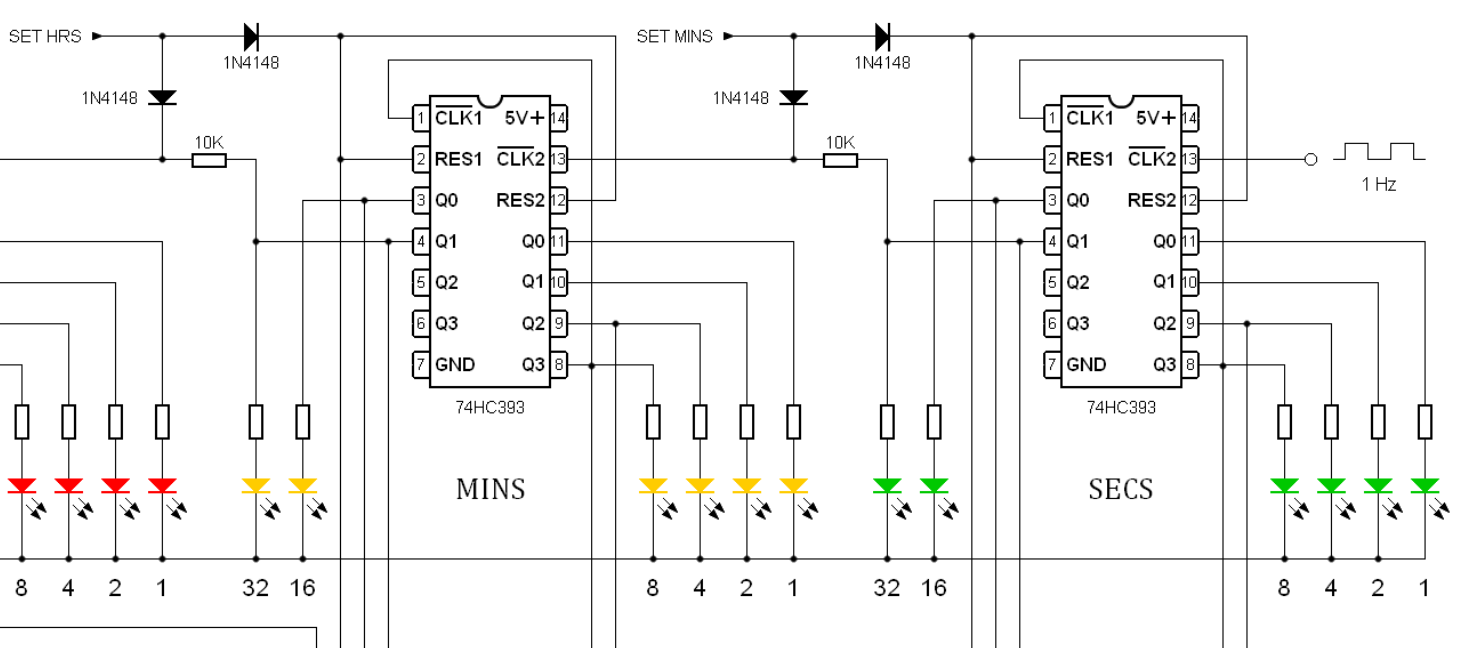}
  \caption{Setting the binary clock \cite{Stuart}}
\end{figure}

The set is connected to the clock input of hours as shown in the diagram above. Every time the set button is pressed on the hours part, a HIGH start is produced on the clock input of the hours counter. This is similar to the HIGH from the `32' output of the previous stage when it becomes active. The input is primed but nothing happens. Once the button is released, a LOW state is produced on the clock input and the counter advances by one. The HIGH produced when the set button is pressed is also fed to the reset pins of the previous (minutes) stage to zero its outputs. This is done because the clock input may be LOW if the time showing is early (which is OK) or HIGH due to the `32' priming signal if the time is late (which is not OK). Resetting the outputs of the previous counter to LOW prevents any conflict between the clock pulse from the counter and the `SET' pulse from the push button.  

\begin{figure}[H]
  \centering
  \includegraphics[width=8cm]{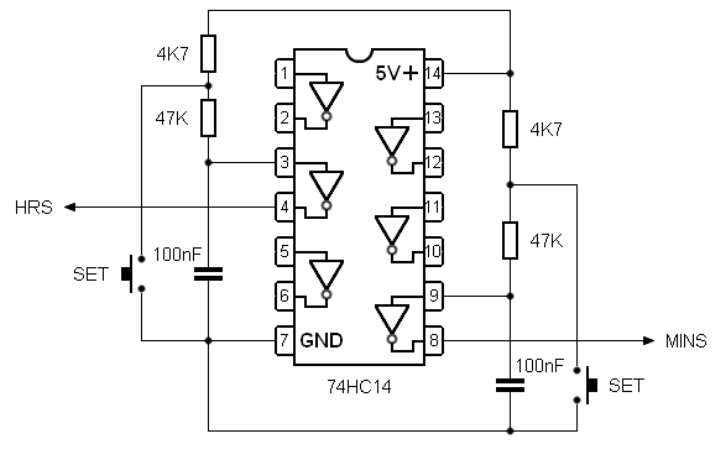}
  \caption{A 74HC14 hex inverting Schmitt trigger IC is used to eliminate switch contact bounce \cite{Stuart}}
\end{figure}

A 74HC14 hex inverting Schmitt trigger IC is used to eliminate switch contact bounce. When a mechnical switch closes, the switch elements will often bounce, before making final contact. This would otherwise produce unwanted clock set pulses in the binary clock circuit. \cite{diodes}

\subsection{1 Hz Clock input}

An Arduino chip was used to produce a 1Hz input for the first stage or the seconds stage of the clock. This is because a clock crystal PCB wasn't available. This increased the costs of the project significantly but we managed to keep it within the specified budget. The following code was used to generate a 1 Hz clock pulse using the ardunio chip:

\begin{lstlisting}[language=C]
void setup()
{
	//initialize digital pin 13 as an output
	pinMode(13,OUTPUT);
}

// the loop function runs over and over again forever
void loop()
{
	digitalwrite(13,HIGH); // turn the LED on (HIGH is the volatage level)
	delay(500); // wait for a second
	digitalwrite(13,LOW); // turn the LED off (LOW is the voltage level)
	delay(500); // wait for a second 
}
\end{lstlisting}

\section{Future Enhancements}

This project could proceed further on to develop a binary clock for smart watches. Smart watches, or \textit{wearable} computers, are digital or computerized wrist watches, with functionality that go beyond time keeping.  Watches such as the Apple \textit{iWatch} and the Samsung \textit{Gear} consists of a small but high-resolution LCD screen to display notifications, calculations, etc. Such companies also provide "Watch" support for an application during development. A smart watch application could be designed to display the time in a \textit{32,16,8,4,2,1,} binary format for tech-savvy enthusiast. Considering that smart watches are usually worn by people who're proficient in technology, the binary clock could be a successfully idea. 

Another extension could be to make the entire watch small enough to fit as a wristwatch. This can be done with the use of smaller ICs and a soldering kit. The aim of this project could be to make a economically priced watch for engineers. This would also require much smaller LEDs and since LEDs have a high light intensity, it could be easily viewable in a low-light environment as compared to regular seven-segment displays on economical digital wrist watch displays that use a cheap back-light to present time.


%

\section*{Acknowledgment}

The author would like to thank his Professor \textit{Yokesh Babu} for helping him decide our project and to obtain the resources he required and also for his invaluable insight in our day to day working of the project.  Furthermore,  I would also like to thank him for his continuous criticism and for setting us on the right path with our project and for the new ideas I received from him.  I would also like to thank our friends who helped us with the Implementation stage of the project and also gave us mental support. Finally we would like to thank \textit{Vellore Institute of Technology} for providing us with such a chance and the required facilities to aid us in our project.

\ifCLASSOPTIONcaptionsoff
  \newpage
\fi



%

%

\begin{IEEEbiography}[{\includegraphics[width=1in,height=1.25in,clip,keepaspectratio]{picture}}]{John Doe}
\blindtext
\end{IEEEbiography}




\end{document}